\documentclass[prl,twocolumn,superscriptaddress,doublespace]{revtex4}%12pt]{article}
\pdfoutput=1
\usepackage{graphicx}
\usepackage{amsmath}
\usepackage{amsfonts}
\usepackage{amssymb}
\usepackage{xcolor, soul}
\usepackage{epstopdf}
\usepackage{float}
\usepackage{subfigure}
\usepackage{hyperref}
\hypersetup{
     colorlinks   = true,
     citecolor    = red,
     linkcolor    = blue,
     urlcolor     = blue,
}
\def\beq{\begin{equation}}
\def\eeq{\end{equation}}
\def\bea{\begin{eqnarray}}
\def\eea{\end{eqnarray}}
\def\be{\begin{equation}}
\def\ee{\end{equation}}
\def\pr{\partial}
\def\nno{\nonumber}
\def\bse{\begin{subequations}}
\def\ese{\end{subequations}}

\graphicspath{{./figs/}}

\begin{document}
%\preprint{APS/123-QED}

\title{Reheating constraints through decaying inflaton}% Force line breaks with \\
%\thanks{}%

\author{Debaprasad Maity}
 \email{debu@iitg.ernet.in}
\affiliation{%
Department of Physics, Indian Institute of Technology Guwahati.\\
 Guwahati, Assam, India 
}%

\date{\today}

\begin{abstract}
In this paper we study the reheating constraints on inflationary models considering perturbatively decaying inflaton. Important difference with the existing analysis is the inclusion of explicit decaying dynamics of inflaton during reheating. One of the important findings of our analysis is the possible existence of maximum reheating temperature $T_{re}^{max}$ considering perturbative reheating scenario. For all the models under consideration the value of this temperature turned out to be $T_{re}^{max} \sim 10^{15}$ GeV. Corresponding to this value of reheating temperature the duration of reheating period assumes naturally small value, $N_{re} \simeq 0.3$ which indicates instantaneous reheating. Most importantly, maximum reheating temperature $T^{max}_{re}$ also leads to a maximum values of scalar spectral index $n^{max}_s$ and inflationary e-folding number  $N^{max}_{cmb}$ in association with the observed CMB scale. Based on our general analysis, we have studied four different inflationary models, and discuss their predictions and constraints.
\end{abstract}
\maketitle
%\tableofcontents

 After the inflation, reheating is the most important phase which sets the initial condition for
 the subsequent standard big-bang evolution of our universe. 
  During this phase all the visible and invisible matter fields are assumed to be produced through the decay of coherently oscillating inflaton field. Therefore, not only inflaton \cite{encyclopedia} but details of its inflationary dynamics is expected to have strong influence on the dynamics of reheating. 
Cosmic Microwave Background (CMB) is crucially dependent upon the background expansion starting from the radiation phase to the present epoch.
Therefore the observed CMB scales and their horizon exit during inflation must be correlated through reheating phase. Reheating\cite{reheating} is the integral part of the inflationary paradigm. However, complicated thermalization process erases many details of this phase. Understanding this phase could answer many unanswered questions 
such as inflationary mechanism itself, baryogenesis, origin of dark matter etc.
A model independent approach has recently been adopted \cite{martin} to
analyze this phase. In this approach the main idea was to parametrize the phase by an effective equation of state $(w_{re})$, and there by constrain it by e-folding number $(N_{re})$ and the reheating temperature $(T_{re})$ through the inflationary parameters. A large class of inflationary models have been studied based on this idea  \cite{martin,reheatingfollows}. Recently there are attempts to consider the explicit decay of inflaton into the analysis \cite{decay}, and subsequently its application to specific inflationary models \cite{decayfollows}. Constraining the reheating temperature through decaying inflation has also be considered \cite{reheatingtemp}. 
However, all the analysis  so far were done based on identifying effective fluid with the reheating equation of state $w_{re}$, and finally express the perturbative reheating temperature in terms of inflaton decay constant. Therefore, the process of decaying inflaton during reheating has not been considered explicitly in the analysis.   
 
As has been mentioned before, our goal in the present paper will be to study the reheating constraints considering the effect of perturbative decay of inflaton into the dynamics of radiation and inflaton. 
Therefore, important parameter of our analysis is the inflaton decay constant. We further assume that the inflaton decays only into the relativistic fields collectively called as radiation.

\textbf {General analysis:}
After inflation, the inflaton oscillates around its minimum and reheating phase starts. Depending upon the coupling parameters, non-perturbative reheating may occur. However, as emphasized before, we ignore non-perturbative reheating for the present purpose. 

We start with the following Einstein's equation for the cosmological scale factor and conservation of energy, 
\bea \label{rhoh}
&&\ddot{n}_{re} = - 2 \dot{n}_{re}^2  + \frac{1-3 w}{6 M_P^2} \rho_{\phi} \\ \nno
&&\dot{\rho}_{\phi} + 3 \dot{n}_{re}(\rho_{\phi}+p_{\phi})
+ \dot{\rho}_{rad} + 4 \dot{n}_{re} \rho_{rad} = 0 .
\eea
Where, "$\rho$"s are the energy densities of two different components.
 At any instant of time during reheating, we parametrize the duration of reheating by e-folding number $n_{re}(t) = \ln (a/a_i)$,
 where "$a$" is the cosmological scale factor. During reheating if we assume the effective equation of state of the inflaton $w =\langle p_{\phi}/\rho_{\phi}\rangle$ to approximately constant, the above conservation equation turned out to be  
\bea \label{rhoh}
\rho_{rad}+ \rho_{\phi} &=&  e^{-4 n_{re}}\left(\rho^{i}_{\phi} + (1-3 w)  \int_{t_i}^{t}  \rho_{\phi} e^{4 n_{re}}  dn_{re}\right).
\eea 
The index "i" stands for the initial stage of reheating, which also marks the end of inflation.
At the beginning of reheating we set $\rho_{rad}(t_i) = 0$. For solving the above set of equations, the boundary condition is $ \dot{n}_{re}(t_i) = \sqrt{{\rho^i_{\phi}}/{3 M_p^2}}$.
 The physical quantity of our interest is the ratio of the radiation energy density and the inflaton energy density. From eq.\ref{rhoh}, one gets
\bea
\frac{\rho^f_{rad}}{\rho^i_{\phi}} = e^{-4 N_{re}} -
\frac {\rho^f_{\phi}} {\rho^i_{\phi}}
+ (1-3 w)e^{-4 N_{re}} \int_i^{f} \frac{\rho_{\phi}}{\rho^i_{\phi}} e^{4 n_r}  dn_{re} ~.
\eea
Where, "f" corresponds to the final value of radiation density. We define total e-folding number during reheating as
$N_{re} = n_{re}(t_f)$.

The main goal is to understand the relation among inflaton's energy density $\rho_{\phi}$, the radiation temperature $T_{rad}$ and the CMB scale $k$.
A particular scale $k$ going out of the horizon during inflation will re-enter the horizon during
usual cosmological evolution. This fact will provide us an important relation among different phases  as follows 
\bea
\ln{\left(\frac {a_k H_k}{a_0 H_0}\right)} =-N_k -N_{re} -
\ln{\left(\frac {a_{re} H_k}{a_0 H_0}\right)},
\label{scalek}
\eea  
where, a particular scale $k$ satisfy the relation $k = a_0 H_0 = a_k H_k$. $(a_{re}, a_0)$, are the cosmological scale factors at the end of reheating phase and at the present time respectively.
$(N_k,H_k)$ are the efolding number and the Hubble parameter respectively during inflation. $H_0$ is the present value of the Hubble constant.

The usual approach so far was to define the effective equation of state of the total energy density during reheating and study its evolution. However, we consider only the radiation part during reheating and try to understand the evolution of its temperature $T_{rad}$ as a function of scalar spectral index associated with  CMB scale.  
The reheating temperature $T_{re}$ is identified with radiation temperature $T_{rad}$
at thermal equilibrium between the decaying inflaton and the radiation. From the entropy conservation of thermal radiation, the relation among $T_{rad}$, and $(T_0, T_{\nu 0} = (4/11)^{1/3} T_0)$, temperature of the CMB photon and neutrino background at the present day respectively, can be written as 
\bea \label{entropy}
g_{re} T_{rad}^3 = \left(\frac {a_0}{a_{re}}\right)^3\left( 2 T_0^3 + 6 \frac 7 8 T_{\nu 0}^3\right).
\eea
Using eq.(\ref{scalek},\ref{entropy}), one arrives at the following well known relation
\bea \label{eqtre}
T_{rad}= \left(\frac{43}{11 g_{re}}\right)^{\frac 1 3}\left(\frac{a_0T_0}{k}\right) H_k e^{-N_k} e^{-N_{re}} ={\cal G}_k e^{-N_{re}} .
\eea
Where, $g_{re} \sim 100$ is the effective number of relativistic degrees of freedom during radiation phase. In our subsequent study we identify the cosmological scale $k$ as the pivot scale set by PLANCK, $k/a_0 = 0.05 Mpc^{-1}$ and compare our result with the corresponding estimated scalar spectral index $n_s = 0.9682 \pm 0.0062$ 
\cite{PLANCK}.

{\bf Exactly solvable case}: In order to understand the possible existence of maximum reheating temperature, we first consider an analytically solvable model where the inflaton is decaying as
\bea
\rho_{\phi}(t) =  \rho^i_{\phi} e^{-3(1+w)n_{re}} e^{-\Gamma n_{re}} .
\eea 
$\Gamma$ is a dimensionless constant,
\begin{figure}[t!]
	\begin{center}
		%\begin{minipage}
		\includegraphics[width=008.0cm,height=03.00cm]{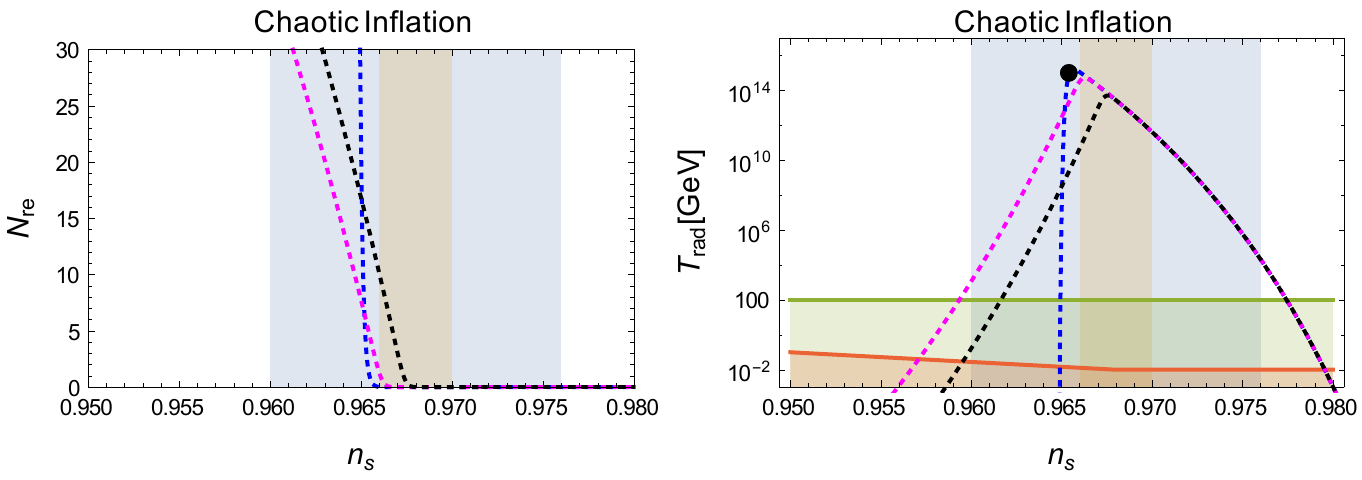}
		\caption{\scriptsize Variation of $(N_{re},T_{rad})$ as a function of $n_s$ have been plotted for $\Gamma = (1.0,0.01, 0.000001)$ corresponding to 		blue, magenta, black curves respectively. The light blue shaded region 
			corresponds to the $1 \sigma$ bounds on $n_s$ from Planck. The brown shaded region corresponds to the $1 \sigma$
			bounds of a further CMB experiment with sensitivity $\pm 10^{-3}$ \cite{limit1,limit2}, using the same central $n_s$ value as
			Planck. Temperatures below the horizontal red line is ruled out by BBN. The deep green shaded region is below the 
			electroweak scale, assumed 100 GeV for reference.} 
		\label{plotexact}
	\end{center}
\end{figure} 
which parametrizes the decay of inflaton. This form of decay essentially modifies the Hubble friction term for the dynamics of inflaton during reheating. With aforementioned ansatz for the decaying inflaton, the solution for radiation density turned out to be 
\bea \label{eqnreexact}
\frac{\rho^f_{rad}}{\rho^i_{\phi}} &=& \frac{\Gamma }{\Gamma  + 3 w -1} \left(e^{-4 N_{re}} -
e^{-3(1+w)N_{re} -\Gamma N_{re}} \right). 
\eea
The second term in the parenthesis is quantifying the fractional amount of inflaton energy
left after the reheating process is over. Expressing  $\rho_f$ in term of radiation temperature as $\rho_{rad}^f=\pi^2 (g_{re}/30) T_{rad}^4$, the eq.\ref{eqnreexact} leads to a maximum radiation temperature for a given $\Gamma$ as   
\bea \label{tmaxexact}
T_{rad}^{max} = \left(\frac{30 \rho^i_{\phi} P}{\pi^2 g_{re}} \right)^{\frac 1 4} \left[x^{\frac {4}{\Gamma +3w-1}}-x^{\frac{3+3w+\Gamma}{\Gamma +3w-1}}
	\right]^{\frac 1 4} ,
\eea
where, $(x = 4/(3+3w+\Gamma), P ={\Gamma }/{(\Gamma  + 3 w -1})$. This also can be clearly seen from the Fig.\ref{plotexact} for each value of $\Gamma$. Now from the perturbative point of view, the value of $\Gamma \leq 1$. However, if we naively extrapolate the above result for large $\Gamma$ most important result turned out to be the existence of a maximum possible temperature, 
\bea
\lim_{\Gamma\gg 1} T_{rad}^{max} =    \left(\frac{30 \rho^i_{\phi}}{\pi^2 g_{re}}\right)^{\frac 1 4} \simeq 2.9 \times 10^{15}~\mbox{GeV}.
\eea
And this value is numerically of the order of same as for limiting perturbative value at $\Gamma=1$ as shown in the figure Fig.\ref{plotexact}.
Therefore, above temperature can be naturally identified as maximum possible reheating temperature. Considering eq.\ref{eqtre}, this also corresponds to a maximum possible value of scalar spectra index $n_s^{max}$ at that temperature. Now identifying associated temperature of the produced radiation in eq.\ref{eqnreexact} with eq.\ref{eqtre}, we arrive at the following exact expression for $(N_{re},T_{rad})$,
\begin{eqnarray} \label{nretre}
&& T_{rad} = {\cal G}_k\left(1-\frac{1}{P}\frac{\pi^2g_{re} {\cal G}_k^4}{3^2.5V_{end}}\right)^{\frac {1}{4-3(1+w)-\Gamma}} . \\
&& N_{re} = \frac {1}{4-3(1+w)-\Gamma}\ln \left[1-\frac {1}{P}\frac{\pi^2g_{re} {\cal G}_k^4}{3^2.5V_{end} }\right]
\label{nretre}
\end{eqnarray}
In the fig.\ref{plotexact}, we have considered three possible values of $\Gamma$ for quadratic inflaton potential. The special value is $\Gamma=1$, for which the equilibrium condition between the inflaton and the radiation can be achieved at the maximum temperature shown as a black dot.
The maximum value of scalar spectral index turned out to be $n_s^{max}\simeq 0.965$. 
This analysis motivates us to subsequently analyze more general cases, and we will show that this conclusion still holds.  

{\bf General case:} 
In this section we will consider general perturbatively decaying inflaton. For this case, 
we express the decaying inflaton as
\bea
\rho_{\phi}(t) =  \rho^i_{\phi} e^{-3(1+w)n_{re}} e^{- \gamma (t-t_i)} ,
\eea 
Where $\gamma$ is effective time independent inflaton decay constant. However, we believe our conclusion will remain same for time dependent $\gamma$, which we will study later. 
Before numerical analysis, let us examine the approximate solution
which has already been discussed in the literature \cite{tmax}.
During the early stage of evolution, approximating
$\rho_{\phi}^i e^{-\gamma t}\simeq \rho_{\phi}^i$, the radiation density can be calculated as
\bea
 \frac{\rho^f_{rad}}{\rho^i_{\phi}} \simeq
 \frac {2\gamma e^{-4 N_{re}}}{(5-3w)\dot{n}_{re}(t_i)}\left(e^{\frac{5-3w}{2}N_{re}} - 1 \right) .
 \eea
As has been discussed for the exactly solvable case in eq.\ref{tmaxexact}, above equation also leads to a maximum radiation temperature \cite{tmax} for a given $\gamma$,
\bea \label{tmax}
T_{rad}^{max} \simeq \left(\frac {39^2 M_p^2  (\dot{n}^i_{re})^2}{(5-3w)^2 \pi^2 g_{re}} \right)^{\frac 1 8} \sqrt{T_{re}} .
\eea
Where, the relation $T_{re}= 0.45 \left({200}/{g_{re}}\right)^{1/4}\sqrt{\gamma M_p}$ has been used. In the same way as we have discussed in the previously discussed exactly solvable case, maximum possible temperature could be obtained, if one identifies a special point where two temperature meets, $T_{rad}^{max} = T_{re}$. 
 Our numerical analysis also shows the maximum reheating temperature at the aforementioned special point, 
\bea
T_{re}^{max} \simeq \left(\frac {39^2  \rho^i_{\phi}}{3(5-3w)^2 \pi^2 g_{re}} \right)^{\frac 1 4}.
\eea
The maximum reheating temperature $T_{re}^{max}$ can also be computed for another exactly solvable case with $w=1/3$. The corresponding result is as follows,   
\bea \label{tmaxw13}
T_{rad}^{max}|_{w=\frac 1 3} &\simeq&  \left(\frac {30 \rho_{\phi}^i  }{ \pi^2 g_{re}} \frac {\gamma}{4 \dot{n}_i + \gamma} \right)^{\frac 1 4} \\\nno
T_{re}^{max}|_{w=\frac 1 3} & =& \lim_{\gamma\gg 4 \dot{n}_i}T_{rad}^{max}|_{w=\frac 1 3} = \left(\frac {30 \rho_{\phi}^i  }{ \pi^2 g_{re}} \right)^{\frac 1 4} .
\eea
This expression is exactly the same as previously discussed  case. For this special value of $w=1/3$, we also have exact expression for parameters $(T_{rad}, N_{re})$ as
\begin{eqnarray} \label{nretrerad}
&& T_{rad} = {\cal G}_k\left(1-\sqrt{\frac{4\rho^i_{\phi}}{3 M_p^2 \gamma^2}} \ln\left[1 - \frac{\pi^2g_{re} {\cal G}_k^4}{3^2.5V_{end} }\right]
 \right)^{-\frac 1 2 }. \\
&& N_{re} = \frac {1}{2}\ln \left[1-\sqrt{\frac{4\rho^i_{\phi}}{3 M_p^2 \gamma^2}} \ln\left[1 - \frac{\pi^2g_{re} {\cal G}_k^4}{3^2.5V_{end} }\right]\right]
\label{nretre}
\end{eqnarray}
\begin{figure}[t!]
	\begin{center}
		%\begin{minipage}
		\includegraphics[width=008.0cm,height=03.00cm]{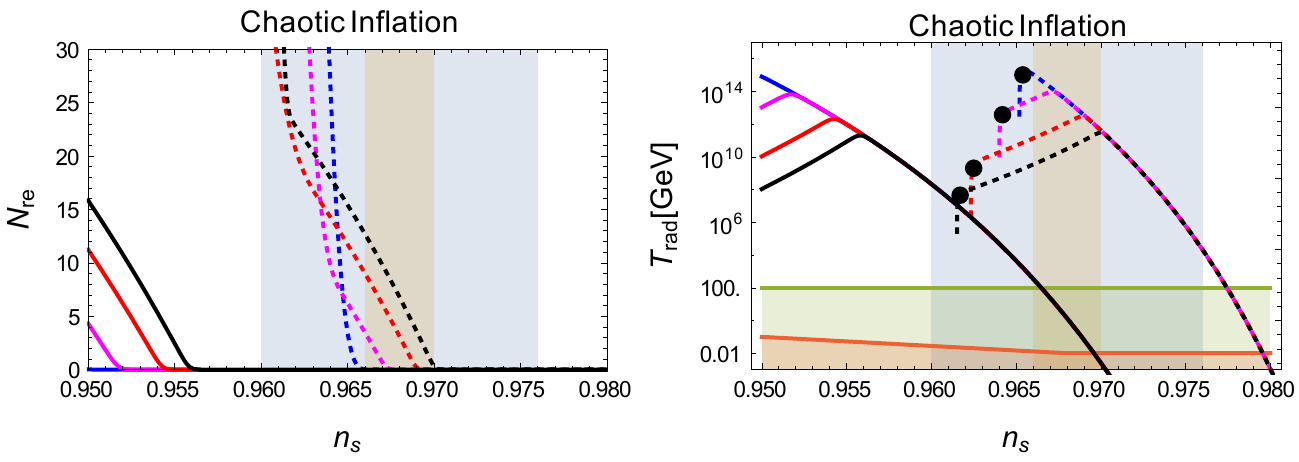}
		\caption{\scriptsize Variation of $(N_{re},T_{rad})$ as a function of $n_s$ have been plotted for four different values of $ \gamma = (10^{13},10^8, 10^2, 10^{-2})  \mbox{GeV}$.
			Blue, magenta, red, black curves correspond to aforementioned four values of $\gamma$ respectively. Various shaded regions are explained in Fig.\ref{plotexact}. Solid and dotted lines correspond to $n=4$ and $n=2$ respectively.} 
		\label{plotchaotic}
	\end{center}
\end{figure}
At this point let us again emphasize the fact that as long as we are in the perturbative regime, the relation among the scalar spectral index $n_s$ and the reheating temperature $T_{re}$ can be understood from our detail analysis above. However, existence of maximum reheating temperature will come if we extrapolate all our formulas for large $\Gamma > \sqrt{2{\rho}_i/(3 M_p^2)}$. For low scale inflation, 
$\Gamma$ could always be in the perturbative regime.
However, for large scale inflation, which we will be considering the system could be in the non-perturbative regime. The detail analysis we will take up for our future publication. To this end, we would like to point out that in the effective reheating equation of state $w_{re}$ description \cite{martin}, the maximum temperature can be explained in the limit of zero reheating e-folding number $N_{re}$. Therefore, large $\Gamma$ limit in our analysis  can be thought of as an equivalent to the zero $N_{re}$ limit of the previously studied reheating constraint studies.

The main point of our is to understand the dynamics of decaying inflaton in the reheating constraint analysis. We think this is the appropriate procedure to understand the relation among $(T_{re}, n_s)$. In the subsequent discussions, we consider different models their predictions.

{\bf Numerical study and constraints:} For general case  let us first describe the strategy of our numerical study. We identify the inflation model dependent input parameters as $N_k, H_k, V_{end}, w$ for a particular CMB scale $k$. Given a canonical inflaton potential $V(\phi)$, the inflationary efolding number $N_k$  and Hubble constant $H_k$ can be expressed as  
\bea
N_k  \simeq \int_{\phi_{end}}^{\phi_k} \frac 3 2 \frac{V(\phi)}{V'(\phi)} d\phi ~~;~~H_k = \frac{1}{3 M_p^2} V(\phi_k),
\eea
were, the field values $(\phi_{end},\phi_k)$ are computed form the condition of end of inflation,
\bea
\epsilon(\phi_{end}) = \frac{1}{2 M_p^2} \left(\frac{V'(\phi_{end})}{V(\phi_{end})}\right)^2 =1, 
\eea
and equating a particular value of scalar spectral index with $n_s(\phi_k)$.
Therefore, we will get explicit relations between $(N_k,n_s^k)$ and $(H_k, n_s^k)$.
The reheating parameters $N_{re},T_{rad}$ will implicitly depend on the scalar spectral index $n_s^k$ for a given scale.

 Associated with each value of $T_{rad}$,
if we assume the radiation domination starts at $T_{rad}$, present CMB scale eq.(\ref{eqtre}), 
fixes the corresponding value of the scalar spectral index. In $T_{rad}~vs~n_s$ plot we have three distinct regions. In the high $n_s$ region, the reheating time parametrized by $N_{rad}$ is significantly small. Therefore, $T_{re}$ increases with the increasing $n_s$ through the specific function ${\cal G}_k$ till the maximum $T^{max}_{rad}$  is reached. In this region, the amount of radiation transfered is significantly small unless we are at the maximum reheating temperature where instant reheating happens. 
 In the intermediate region, increasing $N_{re}$ starts playing role and $T_{rad}$ decreases towards reheating temperature $T_{re}$ where decaying inflaton and radiation  equilibrates. 
We have taken four sample values of decay constants $\gamma = (10^{13},10^8, 10^2, 10^{-2}) \mbox{GeV}$ and corresponding  $T^{max}_{rad}= (1.6\times 10^{15},9.4 \times 10^{13},2.9 \times 10^{12}, 2.9 \times 10^{11}) \mbox{GeV}$ respectively. The aforementioned equilibrium condition fixes the reheating temperature of our universe as $T_{rad} =T_{re}= 0.45 \left({200}/{g_{re}}\right)^{1/4}\sqrt{\gamma M_p}$
Therefore, corresponding perturbative reheating temperatures are $T_{re} =(1.1\times 10^{15},8.4\times10^{12},8.4\times10^{9},8.4\times10^{7})$ GeV. We marked those points as black dots in fig.\ref{plotchaotic}.
In addition for a given $\gamma$ inflaton-radiation equilibrium state during reheating in association with the present CMB scale corresponds to a specific value of inflationary scalar spectral. At the equilibrium temperature perturbative decay of inflaton is expected to be maximum and reheating is assumed to be completed. As expected after this the radiation density falls very fast, therefore, temperature of the radiation will also fall very fast as we decrease $n_s$.  
 
 However, the most interesting point of our analysis is the existence of maximum temperature in eq.{\ref{tmax}. This is clearly seen from all the plots for different models under consideration. At this temperature, the intermediate region shrinks to zero at $T^{max}_{rad} = T_{re}$. The interplay between the observed CMB scale and the dynamics of inflation limit the aforementioned $T^{max}_{rad} = T^{max}_{re} \simeq 10^{15} \mbox{GeV}$ for all the models we have considered. It is also interesting that associated with this $T_{re}^{max}$, eq.\ref{eqtre} and eq.\ref{tmax} predicts maximum value of scalar spectral index $n_s = n_s^{max}$ and inflationary e-folding number $N^{max}_{cmb}$ associated with the CMB scale. 
Therefore, $N_{cmb}^{max}$ is the maximum possible reheating e-folding number required to explain the observed CMB scale assuming the perturbative reheating. In the following discussions we consider various models, and study their predictions and constraints.

\vskip 0.2cm
\textbf {\textit {Chaotic }}\cite{chaotic}\textbf{\textit{inflation}}

As has been emphasized, for the canonical inflationary models, potential will contain all the information. For usual chaotic inflation the potential looks like, 
\bea
V(\phi)= \frac 1 2 m^{4-n} \phi^n.
\eea
Where $n = 2,4,6 \dots$. If we consider only the absolute value of the field, $n=3,5,\dots$ can also be included. The average equation of state of the inflaton during reheating is taken as $w = (n-2)/(n+2)$. As has been mentioned before, the black dots in Fig.\ref{plotchaotic}, correspond to reheating temperature for different value of $\gamma$. 

For $n=2$ case, within the blue shaded ($1 \sigma$) region of $n_s$, the combined effect of CMB measurement and the inflation provide the following limit $0.960 < n_s < 0.9656$, and $ 2\times 10^5 < T_{re} < T_{re}^{max} \simeq 10^{15} $ in GeV unit. Therefore, for $n=2$ chaotic inflation, the maximum value of scalar spectral index $n_s^{max} = 0.9656$. It is also clear from the figure that for $n=4$ or in other word if the inflaton field behaves like a radiation during reheating, prediction of $n_s^{max} \simeq 0.946$ is well outside the observed CMB limit $n_s = 0.9682 \pm 0.0062$ as seen in Fig.\ref{plotchaotic}. Therefore, all models with $n=4,6,\dots$ are strongly disfavored considering the perturbative reheating scenario. 

\begin{figure}
	\begin{center}
		%\begin{minipage}
		\includegraphics[width=008.0cm,height=03.00cm]{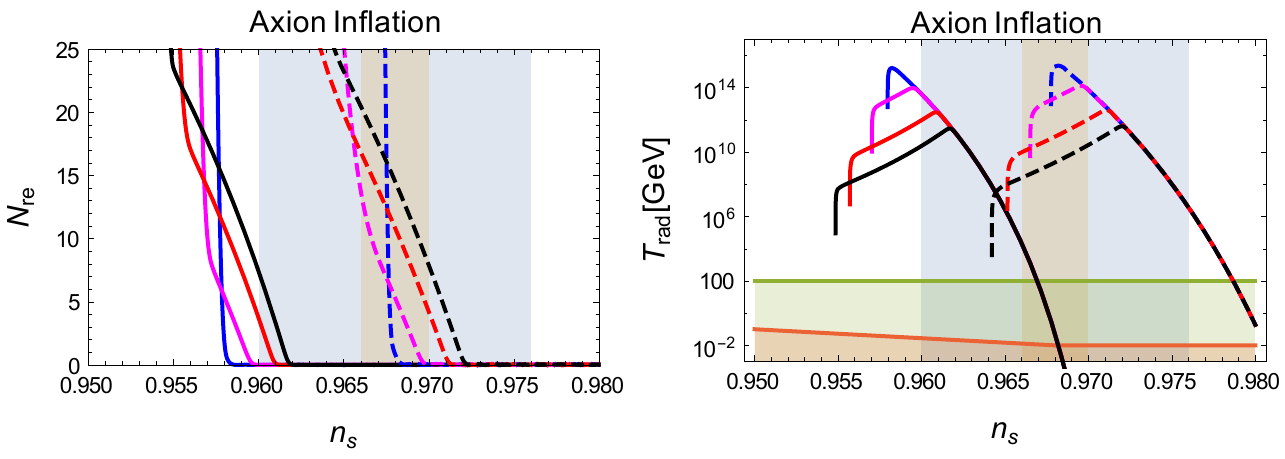}
		\caption{\scriptsize Variation of $(N_{re},T_{rad})$ as a function of $n_s$ have been plotted for $f = 6 M_p$ as solid lines and for $f= 50 M_p$ as dotted lines. Here also we chosen four different values of $ \gamma = (10^{13},10^8, 10^2, 10^{-2})  \mbox{GeV}$. Color codes are same as in Fig.\ref{plotchaotic}.} 
		\label{plotnatural}
	\end{center}
\end{figure}
 
  \begin{figure}[t!]
  	\begin{center}
  		%\begin{minipage}
  		\includegraphics[width=008.0cm,height=03.0cm]{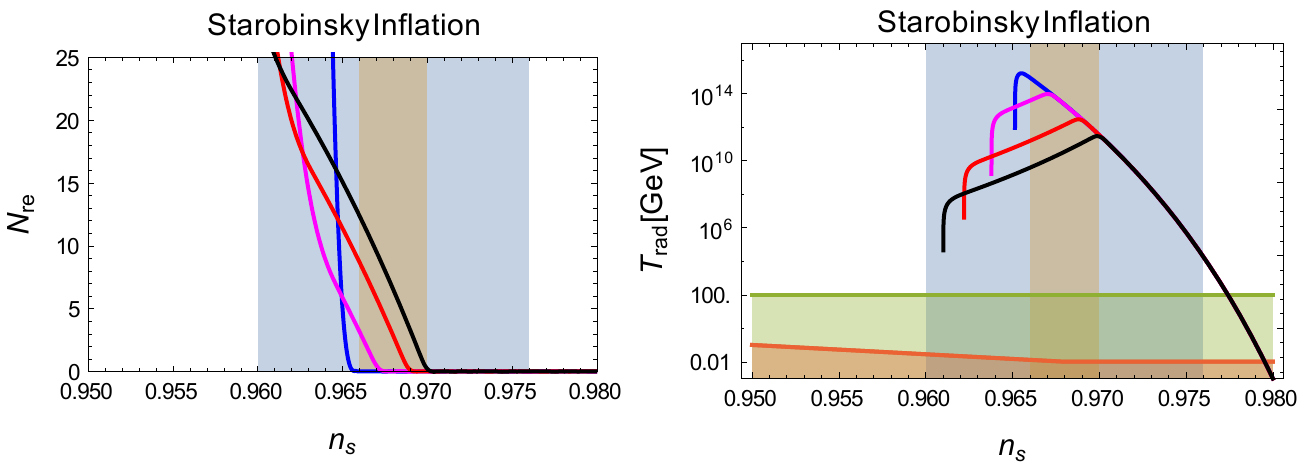}
  		\caption{\scriptsize Variation of $(N_{re},T_{rad})$ as a function of $n_s$ have been plotted. Same values of $\gamma$ and corresponding color code have been considered.} 
  		\label{plotstarobinsky}
  	\end{center}
  \end{figure} 
  
 \vskip 0.2cm
\textbf {\textit {Axion}}\cite{axion}\textbf{\textit{inflation}}

The potential for the axion/natural inflation is 
\bea
V(\phi)=  \Lambda^4 \left[1 + \cos\left(\frac{\phi}{f}\right) \right] .
\eea
where, $(\Lambda,f)$ are the scale of inflation and axion decay constant respectively. The CMB normalization fixes the over all scale of inflation $\Lambda$. Therefore, by tuning $f$ the model can be made compatible with observation as seen in Fig.\ref{plotnatural}. Because of quadratic form of the axion potential near the minimum, we chose $w = 0$ in our analysis. The over all behavior of the $(N_{re}, T_{rad})$ in terms of $n_s$ is similar as for the chaotic models. We have chosen two sample values of axion decay constant $f=(6,10) M_p$. $f=6 M_p$ is disfavored as it predicts maximum value of $n_s^{max}\simeq 0.957$ which is outside the $1\sigma$ region. However for $f =50 M_p$, we have $0.960 < n_s < n_s^{max}\simeq 0.9682$, and $ 2\times 10^2 < T_{re} < T_{re}^{max} \simeq 10^{15} $ in GeV unit. Interestingly, $n_s^{max} = 0.9682$ turned out to be equal to the central value of $n_s$ observed in CMB. 
Therefore, larger value of axion decay constant is favored for the axion model.

\vskip 0.2cm
\textbf {\textit {Starobinski}}\cite{starobinsky}/\textbf{\textit {Higgs}}\cite{higgs}\textbf{\textit{inflation}}

Both of the inflation models are fundamentally different in their non-canonical form. 
However, after the canonical normalization, both the models transform into usual scalar tensor model with the same form of potential as follows,
\bea \label{starohiggs}
V(\phi)=\beta \left(1 - e^{-\sqrt{\frac{2}{3}} \frac {\phi}{M_p}} \right)^2 ,
\eea
where the dimension full parameter $\beta$ takes the following forms,
\bea
\beta_{S} = \frac {1} {4 \alpha}~~~;~~~
\beta_{H} = \frac  {\lambda M_p^4}{\xi^2} .
\eea
Prefixes, $S,H$ stand for Starobinsky and Higgs model respectively.
The aforementioned coupling parameters appear in the non-canonical Lagrangian are as follows,  
\bea
&&{\cal L}_{S} = \frac{M_p^2}{2} R_J(1 + \alpha R_J) + \dots \\
&&{\cal L}_{H} =\frac{M_p^2}{2} R_J + 
\frac{2\xi R_J}{M_p^2}h^2 -\frac 1 2 \pr_{\mu}h \pr^{\mu}h - \frac {\lambda}{4} h^4+ \dots , \nno
\eea 
where, $R_J$ is the Ricci scalar in the Jordan frame.
For the Higgs inflation model one assumes $(\xi > 1, h/M_p > 1)$ during inflation. The inflaton degree of freedom $\phi$ in the  eq.\ref{starohiggs}, are expressed as,
\bea
\phi_{S} = \sqrt{\frac{2}{3}} \ln \left(1+ 2 \alpha R_J \right) ~;~
\phi_{H} = \sqrt{\frac{2 }{3}} \ln \left(1+ \frac{\xi h^2} {M_p^2} \right), \nno
\eea 
in unit of $M_p$. For these system we again consider the equation of state $w=0$. In both these models once we fix the CMB normalization, there is no free parameter to control. Therefore, predictions for both the models are same and quantitatively similar to the chaotic inflation.
From the Fig.\ref{plotstarobinsky}, we clearly see
the combined effect of CMB measurement and the inflation puts following limit $0.960 < n_s < n_s^{max}= 0.9655$, and $ 2\times 10^5 < T_{re} < T_{re}^{max} \simeq 10^{15} $ in GeV unit. In the table \ref{table1}, we summarize all our important predictions for the models we have studied.
\begin{table}
	\caption{Models and their predictions}
	\begin{tabular}{|c|c|c|c|}
		\hline
		\hline
	     & Chaotic & Axion$(f=50M_p)$& Starobinsky/Higgs \\
		\hline
		$T_{re}^{max}$  &$1.60 \times 10^{15}$ & $2.20 \times 10^{15}$ & $1.58 \times 10^{15}$\\
		\hline
		$n_s^{max}$ & 0.96560 &0.9682&0.96548 \\
		\hline
		$N^{max}_{cmb}$ &57.6 &57.2& 55.8\\
				\hline
				
					$N_{re}$ &0.30 &0.32& 0.34\\
					\hline
	\end{tabular}
	\label{table1}
\end{table}

{\bf Summary and discussion:}
One of the main assumptions of our analysis is perturbative decay of inflaton. We have considered two different types of perturbative 
decay processes. Irrespective of the different decay processes, the important out come of our analysis is the existence of maximum reheating temperature $T_{re}^{max} \simeq 10^{15} $ GeV which is nearly independent of the models we have studied. This universality is because of slow roll condition which sets the almost same boundary condition for all the models under consideration. In association with the maximum temperature, the relation between the observed CMB scale and its horizon exit during inflation yields a maximum possible scalar spectral index and e-folding number $(n_s^{max}, N_{cmb}^{max})$ during inflation, which we have shown in the table \ref{table1}. Therefore, considering the explicit decay process into the analysis, the reheating has further restricted the model parameters and even exclude some models. 
For example without considering the tensor spectral index, our analysis excludes power law potential form with $n=4,6 \dots$.      

It would be interesting to further generalize our analysis to consider other decay channels. 
Generalizing our analysis for more general class of inflationary models such as recently found $\alpha$-attractor \cite{alpha} could be interesting. One of the assumptions of our analysis is time independent $w$ and $\gamma$. Depending upon particular model of reheating, $\gamma$ can depend on the background time dependent inflaton field. Therefore, most important generalization of our analysis would be to consider time variation of $(w,\gamma)$. We will take up all these questions for our future publications.

  \hspace{0.5cm}

\end{document}